\documentclass[showpacs,amsmath,nofootinbib,amssymb, epsfig]{revtex4}
\usepackage{graphicx} 
\usepackage{dcolumn}
\usepackage{bm}

\usepackage{color}

\usepackage{amssymb}

\begin{document}

\title{Interacting Holographic Dark Energy model in Brans-Dicke cosmology\\
and \\coincidence problem}
\author{F. Felegary}\email{falegari@azaruniv.ac.ir}\author{F. Darabi}\email{f.darabi@azaruniv.ac.ir;
Corresponding-author}   \affiliation{Department of Physics, Azarbaijan Shahid Madani University, Tabriz, 53714-161 Iran}
\author{M. R. Setare}\email{rezakord@ipm.ir}
\affiliation{Department of Science, Campus of Bijar, University of Kurdistan, Bijar , Iran}

\date{\today}

\begin{abstract}
We study the dynamics of interacting holographic dark energy model in Brans-Dicke cosmology for the future event horizon and  the Hubble horizon cut-offs. { We determine the system of first-order differential equations for the future event horizon and Hubble horizon cut-offs,  and obtain the corresponding fixed points,  attractors, repellers and saddle points}. Finally, we investigate the cosmic coincidence problem in this model for the future event horizon and   Hubble horizon cut-offs and find that for both cut-offs and for a variety of Brans-Dicke parameters the coincidence problem is almost
resolved.

\end{abstract}
\vspace{1cm}
\maketitle

\section{Introduction}
The observations of type Ia Super Novae (SNeIa), Cosmic Microwave Background radiation (CMB) and Larg Scale Structure (LSS) propose
that the expansion of universe is currently accelerating \cite{Riess}. The acceleration of universe
indicates that the present universe is dominated by a mysterious form of energy with negative pressure so called dark energy (DE).
The simplest candidate for dark energy is the cosmological constant. The cosmological constant suffers from the cosmic coincidence
and fine-tuning problems \cite{Weinberg,Copeland}. The cosmic coincidence problem expresses that why dark energy density and
matter density are of order unity \cite{Zimdah1}. A probable way to alleviate the cosmic coincidence problem is to suppose that there is an
interaction between dark energy and dark matter. Also, the cosmic coincidence problem can be alleviated by suitable choice for the form of  interaction
between dark energy and dark matter \cite{Zim,Del,Sadjadi}. The nature of dark energy is unknown and mysterious. Therefore, people have proposed various models for dark energy such as: Quintessence, Tachyon \cite{SetareMR}, Ghost
\cite{GHost}, K-essence, Phantom, Quintom, Chaplygin gas and Holographic \cite{Weinberg,Copeland,Padmanabhan}.
Recently, the holographic dark energy (HDE) model based on the holographic principle was suggested by the following form of energy density \cite{Susskind} 
\begin{equation}\label{rho}
\rho_{\Lambda}=3c^{2}M_{P}^{2}L^{-2},
\end{equation}
where $c$,  $M_{P}$ and $L$ are the numerical constant, the reduced Planck mass and the cut-off radius, respectively. 

{ Cohen {\it et al.} have reasoned
that the dark energy should obey the holographic principle and be constrained
by the infrared (IR) cut-off \cite{Cohen}. In the context
of the holographic dark energy model, Li has discussed, three choices
for the IR cut-off as the Hubble horizon, the particle horizon, and the future
event horizon. He has shown that only the future event horizon is capable
of providing the sufficient acceleration for the universe \cite{Li}. However,  Pav\'on
{\it et al.} have then shown that in the interacting holographic dark energy model \cite{IHDE},
the identification of $L$ with the Hubble horizon can also drive
accelerating universe \cite{Pavon}.  }

The scalar-tensor theories of gravity have been widely studied in cosmology
\cite{STG}. The pioneering study on scalar tensor theories was done by Brans and Dicke \cite{Brans}.   The essence of holographic energy density lies
in a dynamical cosmological constant, so a dynamical frame  like Brans-Dicke theory, instead of general relativity, can accommodate the holographic energy density. Therefore it is useful
to investigate the HDE model in the framework of Brans-Dicke theory.
Such studies  have been carried out in \cite{Baner}. The cosmological application of interacting holographic dark energy density in the framework of Brans-Dicke cosmology has also been studied in \cite{Pasq} and \cite{Sheykhi}. In the
first work, the authors have studied the cosmological implications of
 interacting logarithmic entropy-corrected HDE  model in the framework of Brans-Dicke cosmology, where they have taken the average radius
of Ricci scalar curvature as the infrared (IR) cut-off and obtained the equation of state parameter, the deceleration parameter and the evolution of energy density parameter. In the second work, the equation of state and the deceleration parameter of  holographic dark energy in a non-flat universe was obtained. The author has taken the radius of event horizon measured on the sphere of the horizon as  the infrared (IR) cut-off and has  found that the combination of Brans-Dicke field and holographic dark energy can accommodate phantom crossing for the equation of state of noninteracting holographic dark energy, such that in the presence of interaction between dark energy and dark matter, the transition to phantom regime can be more easily accounted for by the
Brans-Dicke field equations, in comparison to the Einstein field equations.

One of the most important problems in cosmology is the coincidence problem. The coincidence problem  arises by the question that  ``Why  the ratio of  dark matter density to dark energy density is  of order unity"? \cite{campo}. The holographic dark energy model is one of the proposed models to solve this problem in Einstein gravity. In this model, it has been proven that 
the interaction of cold dark matter with holographic dark energy  can solve the coincidence problem \cite{Karwan,dpa,Rev}.
{Also, the coincidence problem has been studied by L. Amendola
\cite{Amendola}
where he first proposed that the coincidence problem gets relieved if there is a scaling attractor and the present state of  universe is close to this fixed point. 

{ Latter
on, Karwan has studied the interacting holographic dark energy model, where
the IR cut-off was taken as the Hubble horizon, and found the fixed points and their stability conditions \cite{Karwan}. He
showed that for some narrow range of the parameters of interacting
model, the cosmic evolution can reach the attractor near the present epoch
and the coincidence problem can be alleviated. Therefore, the coincidence problem became an issue of the parameters and not of the initial conditions}.   Moreover, in Ref. \cite{bisabr}   the coincidence problem in holographic dark energy in $f(R)$ gravity has been explained and it has been shown that the Einstein frame representation of $f(R)$ models may resolve the coincidence problem.}

{ To the authors knowledge, the coincidence problem has not received much attention in the context of Interacting  Holographic Dark Energy model in Brans-Dicke cosmology. Motivated by finding a solution to the coincidence problem in the cosmological
context of alternative
gravity theories
other than Einstein gravity, we  investigate the  coincidence problem in the interacting holographic dark energy model
in the context of Brans-Dicke cosmology. In this regard,  we obtain the equation of state parameter (EoS) for the cases of  future event horizon and  Hubble horizon cut-offs.  
Then, we    determine the system of first-order differential equations and obtain the corresponding
fixed points,  attractors, repellers and saddle points,  for  future event horizon and  Hubble horizon cut-offs.  Finally,  we investigate the coincidence problem
for the interacting  holographic dark energy model with the  future event horizon and Hubble horizon cut-offs in Brans-Dicke cosmology. We show that
 for both cut-offs and for a variety of Brans-Dicke parameters, the  coincidence problem is almost resolved. }
 
\section{Interacting Holographic dark energy model for future event horizon
and Hubble horizon
cut-offs in Brans-Dicke cosmology}

The action of Brans-Dicke theory in the canonical form can be written as \cite{Jordan,Brans,arik}
\begin{equation}
S=\int d^{4}x \sqrt{g}\Big(-\frac{1}{8\omega_{BD}}\phi^{2} R+ \frac{1}{2}g^{\mu\nu}\partial_{\mu}\phi
\partial_{\nu}\phi+L_{M}\Big),
\end{equation}
where $g$, $\omega_{BD}$, $R$ and $L_{M}$ are the determinant of the tensor metric $g^{\mu\nu}$, the Brans-Dicke parameter, the Ricci scalar curvature and $L_{M}$ the lagrangian of the matter, respectively. { Here,
the usual Einstein-Hilbert term $R/G$ has been replaced by the non-minimal coupling term $\phi^{2} R$ so that  $8\omega_{BD}/\phi^{2} $   defines the effective gravitational constant.}
 Variation of the action with respect to the metric $g^{\mu\nu}$ and the Brans-Dicke scalar field $\phi$ gives
\begin{eqnarray}
\phi G_{\mu\nu}=-8\pi T^{M}_{\mu\nu}-\frac{\omega_{BD}}{\phi}\Big(\phi_{,\mu}\phi_{,\nu}-\frac{1}{2}g_{\mu\nu}\phi_{,\kappa}\phi_{,}^{\kappa}
-\phi_{;\mu;\nu}+\Box\phi g_{\mu\nu}\Big),\label{G}
\end{eqnarray}
\begin{equation}
\Box \phi=\frac{8\pi}{2\omega_{BD}+3}T^{M\kappa}_{\kappa},\label{box}
\end{equation}
{where $T^{M}_{\mu\nu}$ is the energy-momentum tensor corresponding to the pressureless dark matter, dark energy, radiation and baryons}. The Friedman- Robertson- Walker (FRW) universe is represented by the line element
\begin{equation}
ds^{2}=dt^{2}-a(t)^{2}\Big(\frac{dr^{2}}{1-kr^{2}}+r^{2}d\Omega^{2}\Big),\label{frw}
\end{equation}
where $a(t)$ is the scale factor and $k$ is the curvature parameter. Now, using Eq. (\ref{frw}) and inserting it in Eqs. (\ref{G}) and (\ref{box}), one can write the Friedmann equation
\begin{equation}
\frac{3}{4\omega_{BD}}\phi^{2}\Big(H^{2}+\frac{k}{a^{2}}\Big)-\frac{1}{2}\dot{\phi}^{2}+\frac{3}{2\omega_{BD}}
H\dot{\phi}\phi=\rho_{m}+\rho_{\Lambda}+\rho_{r}+\rho_{b},\label{frid}
\end{equation}
where $H=\dot{a}/{a}$ is the Hubble parameter. Also $\rho_{m}$, $\rho_{\Lambda}$, $\rho_{r}$ and $\rho_{b}$ are
the pressureless dark matter density,  dark energy density, radiation density   and baryons density, respectively.

We assume that there is an interaction between dark matter and holographic model of dark energy as follows
\begin{equation}
\dot{\rho_{\Lambda}}+3H(1+\omega_{\Lambda})\rho_{\Lambda}=-Q,\label{dotrho}
\end{equation}
\begin{equation}
\dot{\rho_{m}}+3H\rho_{m}=Q,\label{QM}
\end{equation}
and the interaction term is given as follows \cite{Karwan}
\begin{equation}
Q=3H(\lambda_{\Lambda}\rho_{\Lambda}+\lambda_{m}\rho_{m}),\label{QQQ}
\end{equation}
where $\lambda_{\Lambda}$ and $\lambda_{m}$ are the coupling constants.
Also, we suppose that radiation and baryons have no interaction with holographic model of dark energy, so that they obey the continuity equations
\begin{equation}
\dot{\rho_{r}}+4H\rho_{r}=0,\label{Qr}
\end{equation}
\begin{equation}
\dot{\rho_{b}}+3H\rho_{b}=0.\label{Qb}
\end{equation}
{ Here, we may explain about the structure of   system of equations described
so far. The cosmological system is described by two Friedmann equations given by $(0,0)$ and $(1,1)$ components in (\ref{G}), and one wave equation
for the scalar field  (\ref{box}). This wave equation is not an independent
equation because it follows from the Bianchi identities alongside the Friedmann equations (\ref{G}) and conservation equations (\ref{dotrho}), (\ref{QM}). This wave equation is not altered by the interaction between (\ref{dotrho}) and (\ref{QM}) since although the matter and dark energy components do not conserve separately, however the overall fluid (matter plus dark energy) does. We also have the holographic dark
energy (\ref{rho11}). Therefore, our system of equations is not closed and we still have freedom to choose one field in terms of another one. At this point, we may assume that Brans-Dicke scalar field can be described as a power law of the scale factor as $\phi=a^{n}$ \cite{J}. In principle there is no compelling reason for this
choice. However, we will see that this choice with small $|n|$ can lead to consistent results which may justify this specific choice among other possible
choices \cite{J}. }
Thus, one can write
\begin{equation}
\dot{\phi}=nH\phi~~~~~~~,~~~~~~~\ddot{\phi}=(n^{2}H^{2}+n\dot{H})\phi.\label{scale}
\end{equation}
The fractional energy densities are presented by
\begin{equation}
\Omega_{m}=\frac{4\omega_{BD}\rho_{m}}{3\phi^{2}H^{2}},\label{Omegam}
\end{equation}
\begin{equation}
\Omega_{\Lambda}=\frac{4\omega_{BD}\rho_{\Lambda}}{3\phi^{2}H^{2}},\label{Omegalambda}
\end{equation}
\begin{equation}
\Omega_{r}=\frac{4\omega_{BD}\rho_{r}}{3\phi^{2}H^{2}},\label{Omegar}
\end{equation}
\begin{equation}
\Omega_{b}=\frac{4\omega_{BD}\rho_{b}}{3\phi^{2}H^{2}},\label{Omegab}
\end{equation}
\begin{equation}
\Omega_{k}=\frac{k}{a^{2}H^{2}}.\label{Omegak}
\end{equation}
Using Eqs. (\ref{scale}), (\ref{Omegam}), (\ref{Omegalambda}), (\ref{Omegar}), (\ref{Omegab}) and (\ref{Omegak})
and inserting in Eq. (\ref{frid}), we can rewrite the Friedmann equation as follows
\begin{equation}
1+\Omega_{k}-\frac{2}{3}n^{2}\omega_{BD}+2n=\Omega_{m}+\Omega_{\Lambda}+\Omega_{r}+\Omega_{b}.\label{fridman1}
\end{equation}
Also, taking time derivative of Eq. (\ref{frid}) and using
Eqs. (\ref{dotrho}), (\ref{QM}), (\ref{QQQ}), (\ref{Qr}), (\ref{Qb}), (\ref{scale}), (\ref{Omegam}),
 (\ref{Omegalambda}), (\ref{Omegar}), (\ref{Omegab}) and (\ref{Omegak}) we can obtain
\begin{equation}
\frac{\dot{H}}{H^{2}}=\frac{-\frac{9}{4}\Big[(1+\omega_{\Lambda})\Omega_{\Lambda}+\Omega_{m}+\Omega_{b}\Big]
-3\Omega_{r}-\frac{3}{2}\Omega_{k}(n-1)-\frac{3n}{2}+n^{3}\omega_{BD}-3n^{2}}{\frac{3}{2}-n^{2}\omega_{BD}+3n}.\label{Hdot1}
\end{equation}
Moreover, using Eq. (\ref{QQQ}) and inserting in Eq. (\ref{dotrho}), one can obtain the equation of state parameter
\begin{equation}
\omega_{\Lambda}=-1-\frac{\dot{\rho_{\Lambda}}}{3H\rho_{\Lambda}}
-\frac{(\lambda_{\Lambda}\Omega_{\Lambda}+\lambda_{m}\Omega_{m})}{\Omega_{\Lambda}}.\label{eos}
\end{equation}
The future event horizon cut-off is defined by
\begin{equation}
R_{h}=a\int_{t}^{\infty}\frac{dt}{a}=a\int_{x}^{\infty}\frac{dx}{aH}.\label{Rh}
\end{equation}
{where $x=\ln a$ and $a$ is the scale factor.} Now, taking time derivative of Eq. (\ref{Rh}) and using Eq. (\ref{Rh}) one can obtain
\begin{equation}
\dot{R_{h}}=HR_{h}-1.\label{dotRh}
\end{equation}
By considering $L=R_{h}$ and inserting in Eq.  (\ref{rho}), we can obtain the density of 
holographic dark energy for the future event horizon cut-off as follows
\begin{equation}
\rho_{\Lambda}=\frac{3c^{2}\phi^{2}}{4\omega_{BD}R_{h}^{2}}.\label{rho11}
\end{equation}
Using Eq. (\ref{rho11}) and inserting in Eq. (\ref{Omegalambda}) we can write
\begin{equation}
\Omega_{\Lambda}=\frac{c^{2}}{H^{2}R_{h}^{2}}.\label{OOmega}
\end{equation}
Now, taking time derivative of Eq. (\ref{rho11}) and using
Eqs. (\ref{dotRh}), (\ref{rho11}) and (\ref{OOmega}) and inserting in Eq. (\ref{eos})
we can obtain the equation of state parameter of the
holographic dark energy model   for the future event horizon cut-off as follows
\begin{equation}
\omega_{\Lambda}=-\frac{1}{3}-\frac{2n}{3}-\frac{2\sqrt{\Omega_{\Lambda}}}{3c}
-\frac{(\lambda_{\Lambda}\Omega_{\Lambda}+\lambda_{m}\Omega_{m})}{\Omega_{\Lambda}}.\label{eos11}
\end{equation}
Also, the Hubble horizon cut-off is considered as
\begin{equation}
L=H^{-1}.\label{hub}
\end{equation}
Using Eq. (\ref{hub}) and inserting in Eq.  (\ref{rho}), we can obtain the density of 
holographic dark energy for   the Hubble horizon cut-off as follows
\begin{equation}
\rho_{\Lambda}=\frac{3c^{2}\phi^{2}H^{2}}{4\omega_{BD}}.\label{rho33}
\end{equation}
Using Eq. (\ref{rho33}) and inserting in Eq. (\ref{Omegalambda}) we can write
\begin{equation}
\Omega_{\Lambda}=c^{2}.\label{OOmega22}
\end{equation}
Now, taking time derivative of Eq. (\ref{rho33}) and using
Eqs. (\ref{rho33}), (\ref{OOmega22}) and inserting in Eq. (\ref{eos})
we can obtain the equation of state parameter of 
holographic dark energy model for the Hubble horizon cut-off as follows
\begin{eqnarray}\label{eossss}
\omega_{\Lambda}=\Big[-1-\frac{2}{3}
\Big(n+\frac{9-6n^{2}\omega_{BD}+24n+3\Omega_{k}(1+2n)-4n^{3}\omega_{BD}+12n^{2}}{6+12n-4n^{3}\omega_{BD}}\Big)
\nonumber\\+\frac{18-12n^{2}\omega_{BD}+48n+6\Omega_{k}(1+2n)-8n^{3}\omega_{BD}+24n^{2})}{9+18n-6n^{3}\omega_{BD}}
-\frac{(\lambda_{\Lambda}\Omega_{\Lambda}+\lambda_{m}\Omega_{m})}{\Omega_{\Lambda}}\Big]\nonumber\\
\Big[1-\Big(\frac{6c^{2}}{6+12n-4n^{3}\omega_{BD}}\Big)
\Big]^{-1}.
\end{eqnarray}
{Note that if $n=0$, the Brans-Dicke scalar
field becomes trivial, and the equation of state parameter of 
holographic dark energy models for both cut-offs reduce to
their respective expressions in general relativity.  Demanding for negative
$\omega_{\Lambda}$ for justification of accelerating universe, and considering the terms containing the multiplications of $n$ and $\omega_{BD}$, it turns out that
small
values of  $|n|$ can require large values of $\omega_{BD}$, so that the terms
$n^{2}\omega_{BD}$ result in order unity and the terms $n^{3}\omega_{BD}$
become ignorable \cite{J}. Such large values of $\omega_{BD}$ are consistent with
the local astronomical experiments which set a very high lower bound on the Brans-–Dicke parameter  $\omega_{BD}$ \cite{Will}, for instance $\omega_{BD}>10^4$
\cite{Ber}. This consistency may justify our choice of  power law behaviour
of the scale factor as $\phi=a^{n}$, in the previous discussion.}
\section{cosmological dynamics of interacting model in Brans-Dicke cosmology}
{ In this section, we specify the system of first-order differential equations for interacting HDE model with future event horizon and Hubble horizon cut-offs in BD cosmology. Also we  obtain the corresponding fixed points, the attractors, repellers and saddle points.}

\subsection{Future event horizon cut-off}

Taking time derivative of Eq. (\ref{Omegalambda}) and using
Eqs. (\ref{Omegalambda}),  (\ref{Hdot1}), (\ref{rho33}) and $\dot{\Omega_{\Lambda}}=H\acute{\Omega_{\Lambda}}$  can lead to
\begin{eqnarray}
\acute{\Omega_{\Lambda}}=\Omega_{\Lambda}\Big[\Big(1-\frac{1}{HR_{h}}\Big)
\Big(\frac{2c^{2}}{H^{2}R_{h}^{2}\Omega_{\Lambda}}-4\Big)
+\frac{2c^{2}n}{H^{2}R_{h}^{2}\Omega_{\Lambda}}-2n
~~~~~~~~~~~~~~~~~~~~~~~~~~~~~~~~~~\nonumber\\
+\frac{\frac{9}{2}\Big[(1+\omega_{\Lambda})\Omega_{\Lambda}+\Omega_{m}+\Omega_{b}\Big]
+6\Omega_{r}+3\Omega_{k}(n-1)+3n-2n^{3}\omega_{BD}+6n^{2}}{\frac{3}{2}-n^{2}\omega_{BD}+3n}\Big],\label{acute1}
\end{eqnarray}
{ where  $(~ \acute{ }=d/dx)$ and   $( ~\dot{ }=d/dt)$.}  Also, by taking time derivative of Eq. (\ref{Omegam}) and using
Eqs. (\ref{QM}), (\ref{QQQ}), (\ref{scale}), (\ref{Omegam}), (\ref{Omegalambda}), (\ref{Hdot1}) and $\dot{\Omega_{m}}=H\acute{\Omega_{m}}$ we can obtain
\begin{eqnarray}
\acute{\Omega_{m}}=\Omega_{m}\Big(-3+3\lambda_{m}-2n\Big)+3\lambda_{\Lambda}\Omega_{\Lambda}
~~~~~~~~~~~~~~~~~~~~~~~~\nonumber\\
+2\Omega_{m}\Big(
\frac{\frac{9}{4}\Big[(1+\omega_{\Lambda})\Omega_{\Lambda}+\Omega_{m}+\Omega_{b}\Big]
+3\Omega_{r}+\frac{3}{2}\Omega_{k}(n-1)+\frac{3n}{2}-n^{3}\omega_{BD}+3n^{2}}{\frac{3}{2}-n^{2}\omega_{BD}+3n}\Big).\label{acute2}
\end{eqnarray}
Taking time derivative of Eq. (\ref{Omegar}) and using
Eqs. (\ref{Qr}), (\ref{scale}), (\ref{Omegar}), (\ref{Hdot1}) and $\dot{\Omega_{r}}=H\acute{\Omega_{r}}$  can lead to
\begin{eqnarray}
\acute{\Omega_{r}}=-\Omega_{r}(4+2n)
+2\Omega_{r}\Big(
\frac{\frac{9}{4}\Big[(1+\omega_{\Lambda})\Omega_{\Lambda}+\Omega_{m}+\Omega_{b}\Big]
+3\Omega_{r}+\frac{3}{2}\Omega_{k}(n-1)+\frac{3n}{2}-n^{3}\omega_{BD}+3n^{2}}{\frac{3}{2}-n^{2}\omega_{BD}+3n}\Big).\label{acute3}
\end{eqnarray}
Moreover, by taking time derivative of Eq. (\ref{Omegab}) and using
Eqs. (\ref{Qb}), (\ref{scale}), (\ref{Omegab}), (\ref{Hdot1}) and $\dot{\Omega_{b}}=H\acute{\Omega_{b}}$ we can obtain
\begin{eqnarray}
\acute{\Omega_{b}}=-\Omega_{b}(3+2n)
+2\Omega_{b}\Big(
\frac{\frac{9}{4}\Big[(1+\omega_{\Lambda})\Omega_{\Lambda}+\Omega_{m}+\Omega_{b}\Big]
+3\Omega_{r}+\frac{3}{2}\Omega_{k}(n-1)+\frac{3n}{2}-n^{3}\omega_{BD}+3n^{2}}{\frac{3}{2}-n^{2}\omega_{BD}+3n}\Big).\label{acute4}
\end{eqnarray}
Now, using Eqs. (\ref{OOmega}), (\ref{eos11}),  
and inserting in Eqs. (\ref{acute1}),  (\ref{acute2}),  (\ref{acute3}) and  (\ref{acute4}) we obtain
\begin{eqnarray}
\acute{\Omega_{\Lambda}}=\Omega_{\Lambda}\Big[-2+\frac{2\sqrt{\Omega_{\Lambda}}}{c}+
~~~~~~~~~~~~~~~~~~~~~~~~~~~~~~~~~~~~~~~~~~~~~~~~~~~~~~~~~~~~~~~~~~~~~~~~~~~~~~~\nonumber\\
\frac{-\frac{3\Omega_{\Lambda}}{2}-\frac{3\Omega_{\Lambda}^{\frac{3}{2}}}{c}+3n\Omega_{\Lambda}
-\frac{9}{2}(\lambda_{\Lambda}\Omega_{\Lambda}+\lambda_{m}\Omega_{m})+\frac{9}{2}-3n^{2}\omega_{BD}
+12n+\frac{3\Omega_{r}}{2}-2n^{3}\omega_{BD}+6n^{2}}{\frac{3}{2}-n^{2}\omega_{BD}+3n}\Big],\label{acu1}
\end{eqnarray}
\begin{eqnarray}
\acute{\Omega_{m}}=\Omega_{m}\Big[-3+3\lambda_{m}-2n+\frac{3\lambda_{\Lambda}\Omega_{\Lambda}}{\Omega_{m}}+
~~~~~~~~~~~~~~~~~~~~~~~~~~~~~~~~~~~~~~~~~~~~~~~~~~~~~~~~~~~~~~\nonumber\\
\frac{-\frac{9\Omega_{\Lambda}}{2}+3n\Omega_{\Lambda}
-\frac{9}{2}(\lambda_{\Lambda}\Omega_{\Lambda}+\lambda_{m}\Omega_{m})+\frac{9}{2}-3n^{2}\omega_{BD}
+12n+\frac{3\Omega_{r}}{2}-2n^{3}\omega_{BD}+6n^{2}}{\frac{3}{2}-n^{2}\omega_{BD}+3n}\Big],\label{acu2}
\end{eqnarray}
\begin{eqnarray}
\acute{\Omega_{r}}=\Omega_{r}\Big[-4-2n+
~~~~~~~~~~~~~~~~~~~~~~~~~~~~~~~~~~~~~~~~~~~~~~~~~~~~~~~~~~~~~~~~~~~~~~~~~~~~~~~~~~~~~\nonumber\\
\frac{-\frac{3\Omega_{\Lambda}}{2}-\frac{3\Omega_{\Lambda}^{\frac{3}{2}}}{c}+3n\Omega_{\Lambda}
-\frac{9}{2}(\lambda_{\Lambda}\Omega_{\Lambda}+\lambda_{m}\Omega_{m})+\frac{9}{2}-3n^{2}\omega_{BD}
+12n+\frac{3\Omega_{r}}{2}-2n^{3}\omega_{BD}+6n^{2}}{\frac{3}{2}-n^{2}\omega_{BD}+3n}\Big],\label{acu3}
\end{eqnarray}
\begin{eqnarray}
\acute{\Omega_{b}}=\Omega_{b}\Big[-3-2n+
~~~~~~~~~~~~~~~~~~~~~~~~~~~~~~~~~~~~~~~~~~~~~~~~~~~~~~~~~~~~~~~~~~~~~~~~~~~~~~~~~~~~~\nonumber\\
\frac{-\frac{3\Omega_{\Lambda}}{2}-\frac{3\Omega_{\Lambda}^{\frac{3}{2}}}{c}+3n\Omega_{\Lambda}
-\frac{9}{2}(\lambda_{\Lambda}\Omega_{\Lambda}+\lambda_{m}\Omega_{m})+\frac{9}{2}-3n^{2}\omega_{BD}
+12n+\frac{3\Omega_{r}}{2}-2n^{3}\omega_{BD}+6n^{2}}{\frac{3}{2}-n^{2}\omega_{BD}+3n}\Big].\label{acu4}
\end{eqnarray}
Now, we discuss on the dynamical system determined by the Eqs. (\ref{acu1}), (\ref{acu2}), (\ref{acu3}) and (\ref{acu4})
for $\Omega\equiv(\Omega_{\Lambda},\Omega_{m},\Omega_{r},\Omega_{b})$. We solve the dynamical system of equations and obtain
their fixed points by the corresponding matrix of linearization. We can determine
the dynamical character of the fixed points by using the sign of the real part of the eigenvalues.
The real parts of their eigenvalues demonstrate that
the cosmological solutions are attractor, repeller or saddle points \cite{Iakubovskyi}.
When all of the eigenvalues are negative, the fixed point is called an attractor,
when all of the eigenvalues are positive, the fixed point is called a repeller;
otherwise the fixed point is called a saddle point. We present
the eigenvalues of dynamical system  in table 1. Also, in the Eqs. (\ref{acu1}), (\ref{acu2}), (\ref{acu3}) and (\ref{acu4})
we consider $\lambda_{\Lambda}=\lambda_{m}=b^{2}$ \cite{Bin},
where   $\zeta$ and $\varepsilon$ are defined as follows
\begin{equation}
\zeta\equiv\frac{-\frac{9}{2}b^{2}(\Omega_{m}+\Omega_{\Lambda})
+\frac{9}{2}-3n^{2}\omega_{BD}+12n-2n^{3}\omega_{BD}+6n^{2}}{\frac{3}{2}-n^{2}\omega_{BD}+3n},
\end{equation}
\begin{equation}
\chi\equiv\frac{-\frac{9}{2}b^{2}(\Omega_{m}+\Omega_{\Lambda})
+\frac{9}{2}-3n^{2}\omega_{BD}+12n-2n^{3}\omega_{BD}+6n^{2}+\frac{3}{2}\Omega_{r}}{\frac{3}{2}-n^{2}\omega_{BD}+3n}.
\end{equation}
{ Moreover, $\Omega_{m}$ and $\Omega_{\Lambda}$ in the ME model are defined as follows
\begin{equation}
\Omega_{m}=\frac{1}{b^{2}}\Big(0.34-0.55\times 10^{-5}\omega_{BD}\Big),
\end{equation}
\begin{equation}
\Omega_{\Lambda}=1-\frac{2}{3}n^{2}\omega_{BD}+2n-\frac{1}{b^{2}}\Big(0.34-0.55\times 10^{-5}\omega_{BD}\Big),
\end{equation}
and $\Omega_{\Lambda}$, $\Omega_{m}$ and $\Omega_{r}$ in the DMR model are defined as follows
\begin{equation}
\Omega_{m}=\frac{3.03 b^{2}c^{2}\Big(60600-\omega_{BD}\Big)}{(3b^{2}+1)\omega_{BD}-60588-1.81\times 10^{5}b^{2}},
\end{equation}
\begin{equation}
\Omega_{\Lambda}=1.01c^{2},
\end{equation}
\begin{eqnarray}
\Omega_{r}=\frac{1}{(3b^{2}+1)\omega_{BD}-60588-1.81\times 10^{5}b^{2}}\Big(-0.07\omega_{BD}c^{2}b^{2}-1.02 c^{2}\omega_{BD}
+3000c^{2}b^{2}-498 \omega_{BD}^{2}b^{2}+\nonumber\\
6.2\times 10^{4}c^{2}-185\times 10^{-7}\omega_{BD}^{2}+6.02\omega_{BD}b^{2}+2\omega_{BD}
-1.826\times 10^{5}b^{2}-6.09\times 10^{4}\Big).
\end{eqnarray}
According to the above equations, we can see that $\Omega_{\Lambda}$, $\Omega_{m}$ and $\Omega_{r}$ are the fixed points at constant parameters $(c, b^{2},\omega_{BD})$.}
The dominated Dark Matter model is defined as the DM model,
the dominated Baryons model is defined as the B model which shows the early universe, the dominated Radiation model is defined as the R model,
the dominated  interacting dark Matter and dark Energy model is defined as the ME model,
and the dominated  interacting dark Matter and dark Energy model in the presence of radiation model is defined as the DMR model.
In table 2,  we characterize the attractor, repeller and saddle point properties for the fixed points determined in table 1.
\newpage
\hspace{20mm}{\small {\bf Table 1.}} {\small
 Fixed Points and Eigenvalues}\\
    \begin{tabular}{l l l l l p{0.15mm} }
    \hline\hline
  \vspace{0.50mm}
{\footnotesize  $Model$ }&  {\footnotesize ~ $Coordinates$ } & {\footnotesize ~~~~~~~~~~~~~~~~~~~~~~~ $Eigenvalues$ }
\\\hline
\vspace{0.50mm}
{\footnotesize $DM$}&
{\footnotesize $(0,1,0,0)$}&
{\footnotesize $\lambda_{1}=-2+\frac{\frac{9}{2}-3n^{2}\omega_{BD}+12n-2n^{3}\omega_{BD}+6n^{2}}{\frac{3}{2}-n^{2}\omega_{BD}+3n}$}\\
{\footnotesize $$}&{\footnotesize $$}&
{\footnotesize $\lambda_{2}=-3-2n+\frac{\frac{9}{2}-3n^{2}\omega_{BD}+12n-2n^{3}\omega_{BD}+6n^{2}}{\frac{3}{2}-n^{2}\omega_{BD}+3n}$}\\
{\footnotesize $$}&{\footnotesize $$}&
{\footnotesize $\lambda_{3}=-4-2n+\frac{\frac{9}{2}-3n^{2}\omega_{BD}+12n-2n^{3}\omega_{BD}+6n^{2}}{\frac{3}{2}-n^{2}\omega_{BD}+3n}$}\\
\\\hline
\vspace{0.5mm}
{\footnotesize $B$}&
{\footnotesize $(0,0,,0,1)$}&
{\footnotesize $\lambda_{1}=-2+\frac{\frac{9}{2}-3n^{2}\omega_{BD}+12n-2n^{3}\omega_{BD}+6n^{2}}{\frac{3}{2}-n^{2}\omega_{BD}+3n}$}\\
{\footnotesize $$}&{\footnotesize $$}&
{\footnotesize $\lambda_{2}=-3-2n+\frac{\frac{9}{2}-3n^{2}\omega_{BD}+12n-2n^{3}\omega_{BD}+6n^{2}}{\frac{3}{2}-n^{2}\omega_{BD}+3n}$}\\
{\footnotesize $$}&{\footnotesize $$}&
{\footnotesize $\lambda_{3}=-4-2n+\frac{\frac{9}{2}-3n^{2}\omega_{BD}+12n-2n^{3}\omega_{BD}+6n^{2}}{\frac{3}{2}-n^{2}\omega_{BD}+3n}$}\\
\\\hline
\vspace{0.5mm}
{\footnotesize $R$}&
{\footnotesize $(0,0,1,0)$}&
{\footnotesize $\lambda_{1}=-2+\frac{6-3n^{2}\omega_{BD}+12n-2n^{3}\omega_{BD}+6n^{2}}{\frac{3}{2}-n^{2}\omega_{BD}+3n}$}\\
{\footnotesize $$}&{\footnotesize $$}&
{\footnotesize $\lambda_{2}=-3-2n+\frac{6-3n^{2}\omega_{BD}+12n-2n^{3}\omega_{BD}+6n^{2}}{\frac{3}{2}-n^{2}\omega_{BD}+3n}$}\\
{\footnotesize $$}&{\footnotesize $$}&
{\footnotesize $\lambda_{3}=-4-2n+\frac{\frac{15}{2}-3n^{2}\omega_{BD}+12n-2n^{3}\omega_{BD}+6n^{2}}{\frac{3}{2}-n^{2}\omega_{BD}+3n}$}\\
\\\hline
\vspace{0.5mm}
{\footnotesize $ME$}&
{\footnotesize $(\Omega_{\Lambda},\Omega_{m},0,0)$}&
{\footnotesize $\lambda_{1}=-2+\frac{3\sqrt{\Omega_{\Lambda}}}{c}+\frac{3\Omega_{\Lambda}
\Big(-1-\frac{5\sqrt{\Omega_{\Lambda}}}{2c}+2n-\frac{3b^{2}}{2}\Big)}{\frac{3}{2}-n^{2}\omega_{BD}+3n}+\zeta$}\\
{\footnotesize $$}&{\footnotesize $$}&
{\footnotesize $\lambda_{2}=-3-2n+\frac{3\Omega_{\Lambda}
\Big(-\frac{1}{2}-\frac{\sqrt{\Omega_{\Lambda}}}{c}+n+\frac{3b^{2}}{2}\Big)}{\frac{3}{2}-n^{2}\omega_{BD}+3n}+\zeta$}\\
{\footnotesize $$}&{\footnotesize $$}&
{\footnotesize $\lambda_{3}=-4-2n+\frac{3\Omega_{\Lambda}
\Big(-\frac{1}{2}-\frac{\sqrt{\Omega_{\Lambda}}}{c}+n\Big)}{\frac{3}{2}-n^{2}\omega_{BD}+3n}+\zeta$}\\
{\footnotesize $$}&{\footnotesize $$}&
{\footnotesize $\lambda_{4}=-3-2n+\frac{3\Omega_{\Lambda}
\Big(-\frac{1}{2}-\frac{\sqrt{\Omega_{\Lambda}}}{c}+n\Big)}{\frac{3}{2}-n^{2}\omega_{BD}+3n}+\zeta$}\\
\\\hline
\vspace{0.5mm}
{\footnotesize $MER$}&
{\footnotesize $(\Omega_{\Lambda},\Omega_{m},\Omega_{r},0)$}&
{\footnotesize $\lambda_{1}=-2+\frac{3\sqrt{\Omega_{\Lambda}}}{c}+\frac{3\Omega_{\Lambda}
\Big(-1-\frac{5\sqrt{\Omega_{\Lambda}}}{2c}+2n-\frac{3b^{2}}{2}\Big)}{\frac{3}{2}-n^{2}\omega_{BD}+3n}+\chi$}\\
{\footnotesize $$}&{\footnotesize $$}&
{\footnotesize $\lambda_{2}=-3-2n+3b^{2}+\frac{3\Omega_{\Lambda}
\Big(-\frac{1}{2}-\frac{\sqrt{\Omega_{\Lambda}}}{c}+n\Big)-\frac{9b^{2}\Omega_{m}}{2}}{\frac{3}{2}-n^{2}\omega_{BD}+3n}+\chi
$}\\
{\footnotesize $$}&{\footnotesize $$}&
{\footnotesize $\lambda_{3}=-4-2n+\frac{3\Omega_{\Lambda}
\Big(-\frac{1}{2}-\frac{\sqrt{\Omega_{\Lambda}}}{c}+n\Big)-3\Omega_{r}}{\frac{3}{2}-n^{2}\omega_{BD}+3n}+\chi$}\\
{\footnotesize $$}&{\footnotesize $$}&
{\footnotesize $\lambda_{4}=-3-2n+\frac{3\Omega_{\Lambda}
\Big(-\frac{1}{2}-\frac{\sqrt{\Omega_{\Lambda}}}{c}+n\Big)}{\frac{3}{2}-n^{2}\omega_{BD}+3n}+\chi$}\\
\\\hline
 \end{tabular}
 \\ \\ \\ \\
 
\hspace{25mm}{\small {\bf Table 2.}} {\small
 Attractor, Repeller and  Saddle points}\\
    \begin{tabular}{l l l l l p{0.15mm} }
    \hline\hline
  \vspace{0.50mm}
{\footnotesize  $Model $ } & {\footnotesize ~~~~~~~~~ $Repeller$ } &
{\footnotesize~~~~~~  $Attractor$ }  & {\footnotesize~~~~~~~~~~~  $Saddle~ point$ } \\\hline
\vspace{0.5mm}
{\footnotesize $DM$}&
{\footnotesize $~~~~~~~~~~n\geq\frac{1}{2},\lambda_{2}>1$}&
{\footnotesize $~~~~~~~~~~n\leq-\frac{1}{2}$}&
{\footnotesize $~~~~~~~~~~------$}\\
\vspace{0.5mm}
{\footnotesize $B$}&
{\footnotesize $~~~~~~~~~~n\geq\frac{1}{2},\lambda_{2}>1$}&
{\footnotesize $~~~~~~~~~~n\leq-\frac{1}{2}$}&
{\footnotesize $~~~~~~~~~~------$}\\
\vspace{0.5mm}
{\footnotesize $R$}&
{\footnotesize $~~~~~~~~~~n\geq0,n\omega_{BD}\leq3$}&
{\footnotesize $~~~~~~~~~~n\leq-\frac{1}{2},n\omega_{BD}\geq3$}&
{\footnotesize $~~~~~~~~~~------$}\\
\vspace{0.5mm}
{\footnotesize $ME$}&
{\footnotesize $~~~~~~~~~~\lambda_{1},\lambda_{2},\lambda_{3},\lambda_{4}>0$}&
{\footnotesize $~~~~~~~~~~\lambda_{1},\lambda_{2},\lambda_{3},\lambda_{4}<0$}&
{\footnotesize $~~~~~~~~~~------$}\\
\vspace{0.5mm}
{\footnotesize $MER$}&
{\footnotesize $~~~~~~~~~~\lambda_{1},\lambda_{2},\lambda_{3},\lambda_{4}>0$}&
{\footnotesize $~~~~~~~~~~\lambda_{1},\lambda_{2},\lambda_{3},\lambda_{4}<0$}&
{\footnotesize $~~~~~~~~~~------$}\\
\vspace{0.5mm}
\\ \hline
 \end{tabular}
 \vspace{5mm}

\subsection{Hubble horizon cut-off}

Using the calculations similar to those of future event horizon cut-off, we present
the eigenvalues of dynamical system  corresponding to Hubble horizon cut-off
in table 3,
where $\lambda_{\Lambda}=\lambda_{m}=b^{2}$ 
and  $\vartheta$ and $\varpi$ are defined as follows
\begin{eqnarray}
\vartheta=\Big[-1-\frac{2}{3}
\Big(n+\frac{9-6n^{2}\omega_{BD}+24n-4n^{3}\omega_{BD}+12n^{2}}{6+12n-4n^{3}\omega_{BD}}\Big)
\nonumber\\+\frac{18-12n^{2}\omega_{BD}+48n-8n^{3}\omega_{BD}+24n^{2})}{9+18n-6n^{3}\omega_{BD}}
-\frac{b^{2}(c^{2}+1)}{c^{2}}\Big]\nonumber\\
\Big[1-\Big(\frac{6c^{2}}{6+12n-4n^{3}\omega_{BD}}\Big)
\Big]^{-1},
\end{eqnarray}
\begin{eqnarray}
\varpi=\Big[-1-\frac{2}{3}
\Big(n+\frac{9-6n^{2}\omega_{BD}+24n-4n^{3}\omega_{BD}+12n^{2}}{6+12n-4n^{3}\omega_{BD}}\Big)
\nonumber\\+\frac{18-12n^{2}\omega_{BD}+48n-8n^{3}\omega_{BD}+24n^{2})}{9+18n-6n^{3}\omega_{BD}}
-\frac{b^{2}(c^{2}+\Omega_{m})}{c^{2}}\Big]\nonumber\\
\Big[1-\Big(\frac{6c^{2}}{6+12n-4n^{3}\omega_{BD}}\Big)
\Big]^{-1}.
\end{eqnarray}

 Moreover, $\Omega_{m}$ and $\Omega_{r}$ in the MR model are defined as follows
\begin{equation}
\Omega_{m}=-\frac{3b^{2}c^{2}}{3b^{2}+1},
\end{equation}
\begin{eqnarray}
\Omega_{r}=\frac{1}{(36b^{2}+12)(3+6n-3c^{2}-2n^{3}\omega_{BD})}\times ~~~~~~~~~~~~~~~~~~~~~~~~~~~~~~~~~~~~~~~\nonumber\\
\Big\lbrace (1+3b^{2})(-12c^{2}n^{4}\omega_{BD}+16n^{5}\omega_{BD}^{2}+42c^{2}n^{2}\omega_{BD}-48n^{4}\omega_{BD}-72n^{3}\omega_{BD}-24n^{2}\omega_{BD}+
144n^{2}+144n)+\nonumber\\
12c^{2}n^{3}\omega_{BD}(1-3b^{2})+27c^{4}-nc^{2}(162b^{2}+126)+c^{2}(-81b^{2}-63)+108b^{2}+36\Big\rbrace,~~
\end{eqnarray}
and  $\Omega_{m}$ and $\Omega_{b}$ in the MB model are defined as follows
\begin{equation}
\Omega_{m}=-c^{2},
\end{equation}
\begin{eqnarray}
\Omega_{b}=\frac{1}{3b^{2}(3+6n-3c^{2}-2n^{3}\omega_{BD})}\times ~~~~~~~~~~~~~~~~~~~~~~~~~~~~~~~~~~~~~~~\nonumber\\
\Big\lbrace 4n^{2}b^{2}\omega_{BD}(-c^{2}n^{2}+n^{3}\omega_{BD}+c^{2}n+3c^{2}-3n^{2}-4.5n)\nonumber\\
-6c^{2}n^{3}b^{2}\omega_{BD}-18nc^{2}b^{2}+b^{2}(-6n^{2}\omega_{BD}-18c^{2}+36n^{2}+36n+9)+9c^{2}b^{2}\Big\rbrace.
\end{eqnarray}
According to  above equations, we can see that $\Omega_{m}$, $\Omega_{r}$ and $\Omega_{b}$ are the fixed points at constant parameters $(c, b^{2},\omega_{BD})$.
The dominated Dark Matter model is defined as the DM model,
the dominated  dark Matter and Baryons model is defined as the MB model,
and the dominated  dark Matter and  radiation model  is defined as the MR model.
In table 4, we characterize the attractor, repeller and saddle point properties for the fixed points determined in table 3.\\ \\

\hspace{50mm}{\small {\bf Table 3.}} {\small
 Fixed Points and Eigenvalues}\\
    \begin{tabular}{l l l l l p{0.15mm} }
    \hline\hline
  \vspace{0.50mm}
{\footnotesize  $Model$ }&  {\footnotesize ~ $Coordinates$ } & {\footnotesize ~~~~~~~~~~~~~~~~~~~~~~~ $Eigenvalues$ }
\\\hline
\vspace{0.50mm}
{\footnotesize $DM$}&
{\footnotesize $(1,0,0)$}&
{\footnotesize $\lambda_{1}=-3+3b^{2}-2n-\frac{-\frac{9}{2}\Big[(1+\vartheta)c^{2}+1\Big]-3n+2n^{3}\omega_{BD}-6n^{2}}{\frac{3}{2}-n^{2}\omega
_{BD}+3n}-\frac{\frac{9}{2}\Big(b^{2}(1-\frac{6c^{2}}{6+12n-4n^{3}\omega_{BD}})^{-1}-1\Big)}{\frac{3}{2}-n^{2}\omega
_{BD}+3n}$}\\
{\footnotesize $$}&{\footnotesize $$}&
{\footnotesize $\lambda_{2}=-4-2n-\frac{-\frac{9}{2}\Big[(1+\vartheta)c^{2}+1\Big]-3n+2n^{3}\omega_{BD}-6n^{2}}{\frac{3}{2}-n^{2}\omega
_{BD}+3n}$}\\
{\footnotesize $$}&{\footnotesize $$}&
{\footnotesize $\lambda_{3}=-3-2n-\frac{-\frac{9}{2}\Big[(1+\vartheta)c^{2}+1\Big]-3n+2n^{3}\omega_{BD}-6n^{2}}{\frac{3}{2}-n^{2}\omega
_{BD}+3n}$}\\
\\\hline
\vspace{0.5mm}
{\footnotesize $MR$}&
{\footnotesize $(\Omega_{m},\Omega_{r},0)$}&
{\footnotesize $\lambda_{1}=-3+3b^{2}-2n-\frac{-\frac{9}{2}\Big[(1+\varpi)c^{2}+\Omega_{m}\Big]-6\Omega_{r}-3n+2n^{3}\omega_{BD}-6n^{2}}{\frac{3}{2}-n^{2}\omega
_{BD}+3n}-\frac{\frac{9}{2}\Omega_{m}\Big(b^{2}(1-\frac{6c^{2}}{6+12n-4n^{3}\omega_{BD}})^{-1}-1\Big)}{\frac{3}{2}-n^{2}\omega
_{BD}+3n}$}\\
{\footnotesize $$}&{\footnotesize $$}&
{\footnotesize $\lambda_{2}=-4-2n-\frac{-\frac{9}{2}\Big[(1+\varpi)c^{2}+\Omega_{m}\Big]-12\Omega_{r}-3n+2n^{3}\omega_{BD}-6n^{2}}{\frac{3}{2}-n^{2}\omega
_{BD}+3n}$}\\
{\footnotesize $$}&{\footnotesize $$}&
{\footnotesize $\lambda_{3}=-3-2n-\frac{-\frac{9}{2}\Big[(1+\varpi)c^{2}+\Omega_{m}\Big]-6\Omega_{r}-3n+2n^{3}\omega_{BD}-6n^{2}}{\frac{3}{2}-n^{2}\omega
_{BD}+3n}$}\\
\\\hline
\vspace{0.5mm}
{\footnotesize $MB$}&
{\footnotesize $(\Omega_{m},0,\Omega_{b})$}&
{\footnotesize $\lambda_{1}=-3+3b^{2}-2n-\frac{-\frac{9}{2}\Big[(1+\varpi)c^{2}+\Omega_{m}+\Omega_{b}\Big]-3n+2n^{3}\omega_{BD}-6n^{2}}{\frac{3}{2}-n^{2}\omega
_{BD}+3n}-\frac{\frac{9}{2}\Omega_{m}\Big(b^{2}(1-\frac{6c^{2}}{6+12n-4n^{3}\omega_{BD}})^{-1}-1\Big)}{\frac{3}{2}-n^{2}\omega
_{BD}+3n}$}\\
{\footnotesize $$}&{\footnotesize $$}&
{\footnotesize $\lambda_{2}=-4-2n-\frac{-\frac{9}{2}\Big[(1+\varpi)c^{2}+\Omega_{m}+\Omega_{b}\Big]-3n+2n^{3}\omega_{BD}-6n^{2}}{\frac{3}{2}-n^{2}\omega_{BD}+3n}$}\\
{\footnotesize $$}&{\footnotesize $$}&
{\footnotesize $\lambda_{3}=-3-2n-\frac{-\frac{9}{2}\Big[(1+\varpi)c^{2}+\Omega_{m}+\Omega_{b}\Big]-3n+2n^{3}\omega_{BD}-6n^{2}}{\frac{3}{2}-n^{2}\omega_{BD}+3n}+\frac{\frac{9}{2}\Omega_{b}}{\frac{3}{2}-n^{2}\omega_{BD}+3n}$}\\
\vspace{0.5mm}
\\\hline
 \end{tabular}
 \vspace{10mm}

\hspace{15mm}{\small {\bf Table 4.}} {\small
 Attractor, Repeller and  Saddle points}\\
    \begin{tabular}{l l l l l p{0.15mm} }
    \hline\hline
  \vspace{0.50mm}
{\footnotesize  $Model $ } & {\footnotesize ~~~~~~~~~ $Repeller$ } &
{\footnotesize~~~~~~  $Attractor$ }  & {\footnotesize~~~~~~~~~~~  $Saddle~ point$ } \\\hline
\vspace{0.5mm}
{\footnotesize $DM$}&
{\footnotesize $~~~~~~~~~~\lambda_{1},\lambda_{2},\lambda_{3}>0$}&
{\footnotesize $~~~~~~~~~~\lambda_{1},\lambda_{2},\lambda_{3}<0$}&
{\footnotesize $~~~~~~~~~~------$}\\
\vspace{0.5mm}
{\footnotesize $MR$}&
{\footnotesize $~~~~~~~~~~\lambda_{1},\lambda_{2},\lambda_{3}>0$}&
{\footnotesize $~~~~~~~~~~~\lambda_{1},\lambda_{2},\lambda_{3}<0$}&
{\footnotesize $~~~~~~~~~~------$}\\
\vspace{0.5mm}
{\footnotesize $MB$}&
{\footnotesize $~~~~~~~~~~\lambda_{1},\lambda_{2},\lambda_{3}>0$}&
{\footnotesize $~~~~~~~~~~\lambda_{1},\lambda_{2},\lambda_{3}<0$}&
{\footnotesize $~~~~~~~~~~------$}\\
\vspace{0.5mm}
\\ \hline
 \end{tabular}
 \vspace{10mm}

\section{ Coincidence problem for interacting HDE model in Brans-Dicke Cosmology}
\begin{figure}
\centering
{\includegraphics[width=2in]{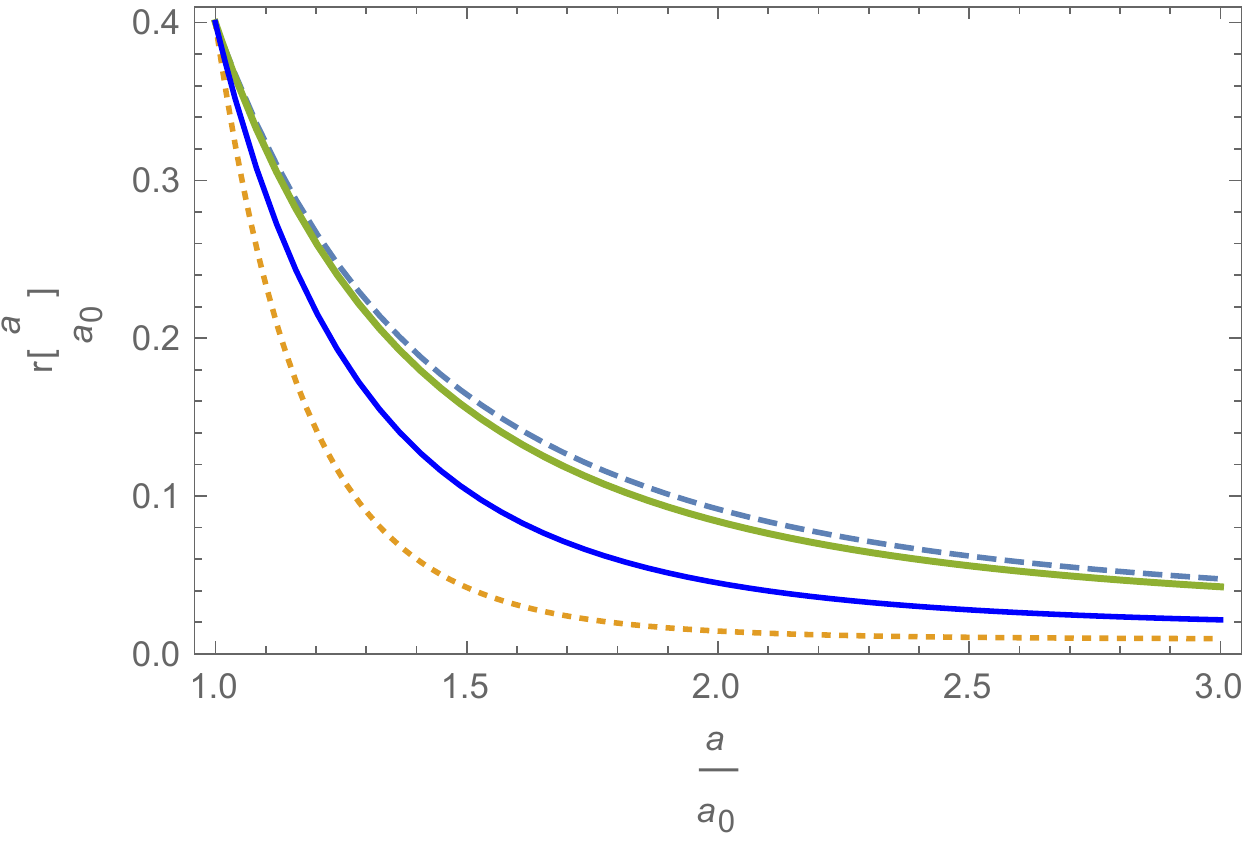}}
\caption{The typical evolution of $r(a/a_0)$ with respect to the scale factor $a$ for the HDE model
with the future event horizon cut-off ($a_0$ denotes for the present value). The dashed line represents  $\omega_{BD}=10000$,
the green line represents  $\omega_{BD}=0$, the black line represents  $\omega_{BD}=-100000$ and
the dotted line represents $\omega_{BD}=-500000$.}
\label{figure1}
\end{figure}

\begin{figure}
\centering
{\includegraphics[width=2in]{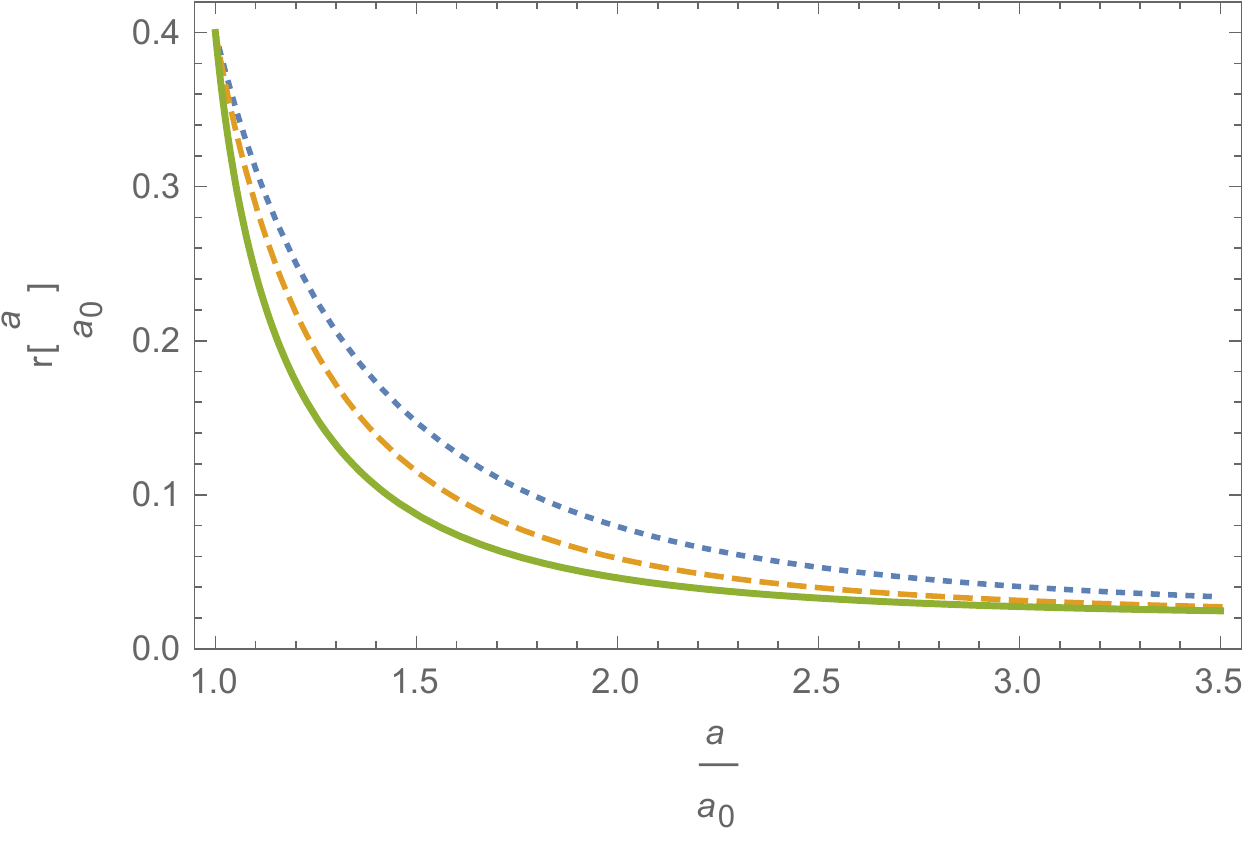}}
\caption{The typical evolution of $r(a/a_0)$ with respect to the scale factor $a$ for the HDE model
with the Hubble horizon cut-off ($a_0$ denotes for the present value). The dotted line represents  $\omega_{BD}=8000$,
the dashed line represents  $\omega_{BD}=100000$, the green line represents  $\omega_{BD}=-50000$.}
\label{figure2}
\end{figure}

In this section, we study the coincidence problem for HDE model in Brans-Dicke cosmology.
We suppose that the contributions of the density of radiation and the density of baryons
are negligible, thus we can write the Friedmann equation as follows
\begin{equation}
\frac{3}{4\omega_{BD}}\phi^{2}\Big(H^{2}+\frac{k}{a^{2}}\Big)-\frac{1}{2}\dot{\phi}^{2}+\frac{3}{2\omega_{BD}}
H\dot{\phi}\phi=\rho_{m}+\rho_{\Lambda}.\label{friddd}
\end{equation}
Using Eqs. (\ref{scale}), (\ref{Omegalambda}) and inserting in Eq. (\ref{friddd}), we obtain
\begin{equation}
\rho_{m}=\frac{3\phi^{2}H^{2}}{4\omega_{BD}}\Big(1+\Omega_{k}-\frac{2}{3}n^{2}\omega_{BD}+2n-\Omega_{\Lambda}\Big).\label{rhmmm}
\end{equation}
Now, we consider the ratio of  density of dark matter to the density of dark energy as follows
\begin{equation}
r=\frac{\rho_{m}}{\rho_{\Lambda}}.\label{r1}
\end{equation}
Using (\ref{Omegalambda}), (\ref{rhmmm}) and inserting in Eq. (\ref{r1}), we can obtain
\begin{equation}
r=\frac{1+\Omega_{k}-\frac{2}{3}n^{2}\omega_{BD}+2n}{\Omega_{\Lambda}}-1.\label{r123}
\end{equation}
Taking time derivative of Eq. (\ref{r123}), using (\ref{dotrho}),(\ref{QM}), (\ref{QQQ}), (\ref{r123}) and
assuming $\lambda_{\Lambda}=\lambda_{m}=b^{2}$ \cite{Bin}, we obtain
\begin{equation}
\dot{r}=3b^{2}H(1+r)^{2}+3H\omega_{\Lambda}r.\label{rrrr}
\end{equation}
{ The advantage of this differential equation is that it includes the Hubble parameter and the interaction between  dark energy and dark matter represented by $b^2$ term. Therefore, knowing the Hubble parameter $H$
and $\omega_{\Lambda}$, one can determine the evolution of $r$}. Using Eq. (\ref{r123}) and inserting in
Eqs. (\ref{eos11}) and (\ref{eossss}), we obtain the equation of state parameters for the future event
horizon and Hubble horizon cut-offs, respectively as
\begin{equation}
\omega_{\Lambda}=-\frac{1}{3}-\frac{2n}{3}-b^{2}(1+r)-\frac{2\sqrt{1+\Omega_{k}-\frac{2}{3}n^{2}\omega_{BD}+2n}}{3c\sqrt{1+r}},\label{eoseos}
\end{equation}
\begin{eqnarray}
\omega_{\Lambda}=\Big[-1-b^{2}(1+r)+\frac{18-12n^{2}\omega_{BD}+48n+6\Omega_{k}(1+2n)
-8n^{3}\omega_{BD}+24n^{2}}{9+18n-6n^{3}\omega_{BD}}~~~~~~~~~~~~~~~~~~~~~~~~~~~~~~~~~~~~~~~~~~~~~~~~~~~~~~~~~~~~~~~~~~~~~\nonumber\\
+\Big(\frac{2c^{2}(1+r)}
{3+3\Omega_{k}-2n^{2}\omega_{BD}+6n}\Big)
\Big(\frac{6n-4n^{4}\omega_{BD}+9-6n^{2}\omega_{BD}
+3\Omega_{k}(1+2n)+48n^{2}-4n^{3}\omega_{BD}}{-6-12n+4n^{3}\omega_{BD}}\Big)\Big]
~~~~~~~~~~~~~~~~~~~~~~~~~~~~~~~~~~~~~~~~~~~~~~~~~~~~~~\nonumber\\
\Big[1-\Big(\frac{6}{6+12n-4n^{3}\omega_{BD}}\Big)
\Big(\frac{6+6\Omega_{k}-4n^{2}\omega_{BD}+12n}{3(1+r)}-c^{2}\Big)\Big]^{-1}.
~~~~~~~~~~~~~~~~~~~~~~~~~~~~~~~~~~~~~~~~~~~~~~\nonumber 
\\~~~~~~~~~~~~~~~~~~~~~~~~~~~~~~~~~~~~~~~~~~~~~~~~~~~~~~~~~~~~~~~~~~~~~~~~~~~~~~~~~~~~~~
~~~~~~~~~~~~~~~~~~~~~~~~~~~~~~~~~~~~~~~~~~~~~~~~~~~~~~~~~~~~~\label{eosssss}
\end{eqnarray}
{  Eq. (\ref{rrrr}) describes the time evolution of $r(a/a_{0})$ and
has not an analytic solution. However, we have plotted numerically the evolution of $r(a/a_{0})$ with
respect to $a/a_{0}$ for the holographic dark energy model with the future event horizon and the Hubble horizon cut-offs. We have plotted Eq. (\ref{rrrr}) for the observational data at  present time.}
Now, for present time we consider $c^{2}=1.1$ \cite{Maa} , $n=0.005$ \cite{Lu}, $\Omega_{k}=0$, $b^{2}=0.02$ \cite{Maa}
and using Eqs. (\ref{eoseos}), (\ref{eosssss}) and inserting in Eq. (\ref{rrrr}), we can plot $r(a/a_0)$ in the figures (\ref{figure1}) and
(\ref{figure2}) for the HDE model with the future event horizon and the Hubble horizon cut-offs.
We find that for both cut-offs and for a variety of Brans-Dicke parameters, the fraction $r(a/a_0)$ experiences a rather fast decreases at small scale factors, which means that at early stages of universe expansion the
dark matter is transformed into dark energy in a rather high rate. { Although
$r$ decreases with scale factor at late times, however
for rather large scale factors the fraction $r(a/a_0)$ approaches  to  almost constant small values, which means
that the interaction between  dark energy and dark
matter is approaching to a frozen phase (without interaction) at late times and  the transfer rate of dark matter
to dark energy is  almost vanishing. 
In other words, unless at infinite scale factor where the fraction $r$ is vanishing,  $r$ is a nonvanishing and almost constant small value at late time,
which may justify that the fraction of dark matter to dark energy will remain
almost constant for a sufficient period of accelerating phase of the universe.
This may alleviate the coincidence problem in our model}.
\section{Concluding remarks}\label{Con}
{ In the interacting and non-interacting HDE models, the future event horizon IR cut-off is the most favored IR cut-off in comparison to the Hubble horizon and particle horizon cut-offs. This is because it can
lead to an accelerated expansion as well as solving the coincidence problem.
The Hubble horizon cut-off is also a viable cut-off just in the interacting HDE models, because it can predict acceleration and solve the coincidence
problem for interacting HDE models. However, the particle horizon cut-off plays no important role in any interacting and non-interacting HDE models, because it is impossible to obtain an accelerated expansion for this case.
In our model of interacting HDE in the Brans-Dicke cosmology, we have studied
both future event horizon and Hubble horizon cut-offs and found that both
models can predict a negative equation of state parameter, necessary for the accelerated expansion, and alleviate the coincidence problem, for a variety of Brans-Dicke parameters. The existence of these viable cut-offs in the
context of HDE model influences drastically the Brans-Dicke cosmology, because
the Brans-Dicke cosmology by itself cannot predict the accelerated expansion
of the universe. However, when combined by HDE model with future event horizon and Hubble horizon cut-offs, it can predict accelerated expansion of the
universe and resolve the coincidence problem, simultaneously. We also determined the system of first-order differential equations and obtained the corresponding
fixed points, attractors, repellers and saddle points for both of future event horizon and Hubble horizon cut-offs. }

{ We may discuss on the new ingredients and significant progresses of this work in comparison to the past related ones.  From
 one hand, the
coincidence problem has already been studied in the interacting
and non-interacting holographic dark energy models in the context of GR and  $f(R)$ and it was found that the coincidence problem can be alleviated for
 both Hubble horizon and future event horizon cut-offs. On the
other hand, the
coincidence problem has already been studied in the non-interacting holographic dark energy model in the context of Brans-Dicke cosmology.
Therefore, the one important ingredient of the present work is the study of coincidence problem in the  interacting
holographic dark energy model in the context of Brans-Dicke cosmology. 
It is obvious that in the absence of HDE and
Brans-Dicke parameter, our model is reduced to the $\Lambda$CDM model in general relativity. In principle, the search for alternative theories of
$\Lambda$CDM model in general relativity is because of coincidence problem
which occurs in the $\Lambda$CDM model in general relativity. Therefore, any condition in the alternative theories, like our model, which can resolve
or alleviate this problem is of particular importance. In this
work, we have found that the Brans-Dicke theory, as the most important alternative theory of general relativity, when combined by the HDE model with future event horizon
and Hubble horizon cut-offs, has the capability of addressing the coincidence problem for a large variety of the Brans-Dicke parameter. In this regard,
we have obtained the differential equation for description of the time evolution of $r(a/a_{0})$  and plotted numerically the evolution of $r(a/a_{0})$ for HDE model with the future event horizon and the Hubble horizon cut-offs, using the observational data at  present time and considering $c^{2}=1.1$, $n=0.005$, $\Omega_{k}=0$, $b^{2}=0.02$.  
We found that the fraction $r(a/a_0)$ experiences a rather fast decreases at small scale factors, which means that at early stages of universe expansion the
dark matter is transformed into dark energy in a rather high rate. Although
$r$ decreases with scale factor at late times, however
for rather large scale factors the fraction $r(a/a_0)$ approaches  to  almost constant small values, which means
that the interaction between  dark energy and dark
matter is approaching to a frozen phase (without interaction) at late times and  the transfer rate of dark matter
to dark energy is  almost vanishing for a large variety of the Brans-Dicke parameter. This alleviates the coincidence problem
 through a more flexible way in comparison to  the previous ones, especially the $\Lambda$CDM model in general relativity.
 }

\end{document}